\documentstyle[aps]{revtex}

\begin{document}

\title{SUBLUMINAL AND SUPERLUMINAL SOLUTIONS IN VACUUM
OF THE MAXWELL EQUATIONS AND THE MASSLESS DIRAC EQUATION}

\author{Waldyr A. Rodrigues, Jr.\thanks{walrod@ime.unicamp.br} and
Jayme Vaz, Jr.\thanks{vaz@ime.unicamp.br}}

\address{Department of Applied Mathematics - IMECC\\
State University at Campinas (UNICAMP)\\
CP 6065, 13081-970, Campinas, SP, Brazil}

\maketitle

\begin{abstract}
We show that Maxwell equations and Dirac equation (with zero mass term) have
both subluminal and superluminal solutions in vacuum. We also discuss the
possible fundamental physical consequences of our results.
\end{abstract}

\section{Introduction}
According to Bosanac \cite{REF-1} there is no formal proof based only on
Maxwell equations that no electromagnetic wave packet can travel
faster than the vacuum speed of light $c$ ($c=1$ in the natural units
to be used here). Well, the main purpose of this paper is to show
that Maxwell equations (and also the Dirac equation with zero mass)
have superluminal solutions $(v>1)$ and also subluminal solutions
$(v<1)$ in vacuum.

This paper is organized as follows. In sect.2 we introduce
some mathematical tools that will be used. In sect.3
we show how to construct the se called subluminal and
superluminal solutions of the free Maxwell equations.
In sect.4 the subluminal and superluminal solutions
of the massless Dirac equation are discussed. Finally
in sect.5 we discuss some of the possible physical
implications of these results.

\section{Mathematical Preliminaries}
In order to discuss these
new solutions of Maxwell and Dirac
equations in an unified way we briefly recall how these equations can
be written in the Clifford and Spin-Clifford bundles over Minkowski
spacetime. Details concerning these theories can be found in
\cite{REF-2,REF-2a,REF-2b}.

Let ${\cal M}=(M,g,D)$ be Minkowski spacetime. $(M,g)$ is a four dimensional
time oriented and space oriented Lorentzian manifold, with $M\simeq
{\sl I \!\! R}^4$ and $g \in {\rm sec}(T^*M \times T^*M)$ being a
Lorentzian metric
of signature (1,3). $T^*M$ [$TM$] is the cotangent [tangent] bundle.
$T^*M = \cup_{x\in
M} T^*_xM$ and $TM = \cup_{x\in M}T_xM$, and $T_xM \simeq T^*_xM \simeq
{\sl I \!\! R}^{1,3}$, where ${\sl I \!\! R}^{1,3}$ is the Minkowski
vector space \cite{REF-3,REF-3a}.
$D$ is the Levi-Civita connetion of $g$, i.e., $Dg=0$,
$\mbox{\boldmath $T$}(D) =0$. Also
$\mbox{\boldmath $R$}(D)=0$, $\mbox{\boldmath $T$}$ and
$\mbox{\boldmath $R$}$ being
respectively the torsion and curvature
tensors. Now, the Clifford bundle of differential forms ${\cal C}\ell(M)$ is
the bundle of algebras ${\cal C}\ell(M) = \cup_{x\in M} {\cal
C}\ell(T^*_xM)$, where
$\forall x\in M, {\cal C}\ell(T^*_xM) = {\cal C}\ell_{1,3}$, the so
called spacetime
algebra \cite{REF-4,REF-4a}. Locally as a linear space over the real field
${\sl I \!\! R},{\cal C}\ell(T^*_x(M)$ is isomorphic to the Cartan
algebra $\bigwedge (T^*_x(M)$ of the
cotangent space  and
$\bigwedge(T^*_x M) = \sum^4_{k=0} \bigwedge {}^k(T^*_x M)$,
where $\bigwedge^k(T^*_x M)$ is the $4 \choose k$ dimensional space of
$k$-forms. The
Cartan bundle $\bigwedge(M) = \cup_{x\in M} \bigwedge(T^*_x M)$ can then be
thought as
``imbedded" in ${\cal C}\ell(M)$. In this way sections of ${\cal
C}\ell(M)$ can be
represented as a sum of inhomogeneous differential forms. Let $\{
e_\mu = \frac{\partial}{\partial x^\mu}\} \in {\rm sec} TM, (\mu =
0,1,2,3)$ be an orthonormal basis
$g(e_\mu, e_\nu) = \eta_{\mu\nu} = {\rm diag}(1,-1,-1,-1)$ and let $\{
\gamma^\nu = d x^\nu \} \in
{\rm sec} \bigwedge^1(M) \subset {\rm sec} {\cal C}\ell(M)$ be the
dual basis. Then, the
fundamental Clifford product (in what follows to be dennoted by
juxtaposition of symbols) is generated by
$\gamma^\mu\gamma^\nu+\gamma^\nu\gamma^\mu = 2\eta^{\mu\nu}$
and if ${\cal C} \in {\rm sec}{\cal C}\ell(M)$ we have
$$
{\cal C} = s+ v_\mu \gamma^\mu + \frac{1}{2!} b_{\mu\nu} \gamma^\mu\gamma^\nu +
\frac{1}{3!} a_{\mu\nu\rho} \gamma^\mu\gamma^\nu\gamma^\rho + p \gamma^5
$$
where $\gamma^5 = \gamma^0\gamma^1\gamma^2\gamma^3
= dx^0 dx^1 dx^2 dx^3$  is the volume element and
$s,v_\mu, b_{\mu v}, a_{\mu\nu\rho}, p\in {\rm sec} \bigwedge^0(M)
\subset {\rm sec}
{\cal C}\ell(M)$. For $A_r \in {\rm sec} \bigwedge^r(M)\subset {\rm sec}
{\cal C}(M), B_s \in {\rm sec}
\bigwedge^s(M)$ we define \cite{REF-4,REF-4a}
$A_r\cdot B_s = \langle A_rB_s \rangle_{|r-s|}$ and
$A_r\wedge B_s = \langle A_rB_s \rangle_{r+s}$,
where $\langle \hspace{1ex} \rangle_k$ is the component in
$\bigwedge^k(M)$ of the Clifford field.

Besides the vector bundle ${\cal C}\ell(M)$ we need
also to introduce another vector
bundle ${\cal C}\ell_{{\rm Spin}_+(1,3)}(M)$
$[{\rm Spin}_+(1,3) \simeq
{\rm SL}(2,{\sl I \!\!\!\! C})]$
called the Spin-Clifford bundle. We can show that
${\cal C}\ell_{{\rm Spin}_+(1,3)}(M)\simeq {\cal C}\ell(M)/{\cal R}$,
i.e., it is a quotient
bundle. This means that sections of ${\cal C}\ell_{{\rm Spin}_+(1,3)}(M)$ are
some special equivalence classes of sections of the Clifford bundle,
i.e, they are equivalence sections of non-homogeneous differential
forms (see eqs.(\ref{eq.1},\ref{eq.2}) below).

Now, as well known an electromagnetic field is represented by $F \in
{\rm sec} \bigwedge^2(M) \subset {\rm sec} {\cal C}\ell(M)$. How to
represent the Dirac spinor
fields in this formalism? We can show that even sections of
${\cal C}\ell_{spin_+(1,3)}(M)$, called Dirac-Hestenes spinor fields,
do the
job. If we fix two orthonormal basis, $\Sigma = \{\gamma^\mu\}$ as before,
and
$\dot{\Sigma} = \{\dot{\gamma}^\mu = R\gamma^\mu \widetilde{R} =
\Lambda^\mu_\nu \gamma^\nu \}$ with $\Lambda^\mu_\nu \in
{\rm SO}_{+}(1,3)$ and
$R\in {\rm Spin}_+(1,3)$, $R\widetilde{R} = \widetilde{R} R =1$,
and where
$\, \widetilde{} \, $ is the reversion operator in ${\cal
C}\ell_{1,3}$ \cite{REF-4,REF-4a}, then the
representations of an even section $\mbox{\boldmath $\psi$} \in {\rm sec}
{\cal C}\ell_{{\rm Spin}_+(1,3)}(M)$ are the sections $\psi_\Sigma$ and
$\psi_{\dot{\Sigma}}$ of ${\cal C}\ell(M)$ related by
\begin{equation}
\label{eq.1}
\psi_{\dot\Sigma} = \psi_\Sigma R
\end{equation}
and
\begin{equation}
\label{eq.2}
\psi_\Sigma = s + \frac{1}{2!} b_{\mu\nu} \gamma^\mu \gamma^\nu + p \gamma^5
\end{equation}
Note that $\psi_{\Sigma}$ has the correct number of degrees of
freedom in order to represent a Dirac spinor field, which is not the
case with the so called Dirac-K\"ahler spinor field.

Let $\star$ be the Hodge star operator $\star :\bigwedge^k(M) \rightarrow
\bigwedge^{4-k}(M)$.
Then we can show that if $A_p \in {\rm sec} \bigwedge^p(M)
\subset {\rm sec}
{\cal C}\ell(M)$ we have
$\star A = \widetilde{A} \gamma^5$.
Let $d$ and $\delta$ be respectively the diferential and Hodge
codifferential operators acting on sections of $\bigwedge(M)$. If
$\omega_p \in {\rm sec}
\bigwedge^p(M)\subset {\rm sec} {\cal C}\ell(M)$, then $\delta \omega_p
= (-)^p \star^{-1} d \star \omega_p$, with $\star^{-1}\star = {\rm
identity}$.

The Dirac operator acting on sections of ${\cal C}\ell(M)$ is the invariant
first order differential operator
\begin{equation}
\partial = \gamma^\mu D_{e_{\mu}} ,
\end{equation}
and we can show the very important result \cite{REF-5}:
\begin{equation}
\partial = \partial \wedge \,  + \, \partial \cdot = d-\delta .
\end{equation}

With these preliminaries we can write Maxwell and Dirac equations as
follows \cite{REF-6}:
\begin{equation}
\partial F = 0 ,
\end{equation}
\begin{equation}
\partial \psi_{\Sigma} \gamma^1\gamma^2 + m \psi_\Sigma \gamma^0 =0 .
\end{equation}
If $m=0$ we have the massless Dirac equation
\begin{equation}
\label{eq.10}
\partial \psi_\Sigma = 0 ,
\end{equation}
which is Weyl's one when $\psi_\Sigma$ is reduced to a Weyl spinor field.
Note that in this formalism Maxwell equations condensed in a single
equation!
Also, the specification of $\psi_\Sigma$ depends on the frame
$\Sigma$.

If one wants to work in terms of the usual spinor formalism,
we can translate our results by choosing, for example,
the standard matrix representation of $\{\gamma^\mu\}$, and
for $\psi_{\Sigma}$ given by eq.(\ref{eq.2}) we have the
following (standard) matrix representation:
\begin{equation}
\psi = \left( \begin{array}{cc}
              \phi_1 & -\phi_2^* \\
	      \phi_2 & \phi_1^*
	      \end{array} \right) ,
\end{equation}
where
\begin{equation}
\phi_1 = \left( \begin{array}{cc}
                s - ib_{12} & b_{13}-ib_{23} \\
		-b_{13}-ib_{23} & s+ib_{12}
		\end{array} \right) , \quad
\phi_2 = \left( \begin{array}{cc}
                -b_{03}+i p & -b_{01}+ib_{02} \\
		-b_{01} - ib_{02} & b_{03}+ i p
		\end{array} \right) .
\end{equation}
Right multiplication by
$$
\left( \begin{array}{c}
          1 \\ 0 \\ 0 \\ 0
       \end{array} \right)
$$
gives the usual Dirac spinor.

Before we present the subluminal and superluminal
solutions $F_<$ and $F_>$ we shall define precisely an inertial reference
frame (irf) \cite{REF-3,REF-3a}. An irf $I\in {\rm sec} TM$ is a timelike
vector
field pointing into the future such that $g(I,I) =1$ and $DI=0$. Each
integral line of $I$ is called an observer. A chart $\langle x^\mu
\rangle$ of
the maximal atlas of $M$ is called naturally adapted to $I$ if
$I=\partial/\partial x^0 $. Putting $I=e_0$, we can find $e_i =
\partial/\partial x^i$ such that
$g(e_\mu,e_\nu) = \eta_{\mu\nu}$ and the coordinates $\langle
x^\mu\rangle$ are the usual Einstein-Lorentz ones and have a
precise operational meaning \cite{REF-7}. $x^0$ is measured by
``ideal clocks" at rest synchronized ``a la Einstein'' and $x^i$
are measured with ``ideal rulers".

\section{Subluminal and Superluminal solutions
of the Maxwell equations}
Let $A \in {\rm sec}
\bigwedge^1(M) \subset {\rm sec} {\cal C}\ell(M)$ be the
vector potential. We
fix the Lorentz gauge, i.e., $\partial \cdot A = - \delta A =0$ such that
$F=\partial A =
(d-\delta)A = dA$. We have the following:

\vspace{1ex}

\noindent {\bf Theorem:} Let $\Pi \in {\rm sec} \bigwedge^2(M)
\subset {\rm sec} {\cal C}\ell(M)$ be the
so called Hertz potential. If $\Pi$
satisfies the wave equation, i.e, $\partial^2 \Pi =
\eta^{\mu\nu}\partial_\mu\partial_\nu \Pi = -(d \delta + \delta d)\Pi
=0$, and if $A =
-\delta\Pi$, then $F = \partial A$
satisfies the
Maxwell equations $\partial F =0$.

\vspace{1ex}

The proof is trivial. Indeed $A = - \delta \Pi$, then $\delta A = - \delta^2
\Pi = 0$ and $F =\partial A =dA$. Now $\partial F = (d-\delta) \
(d-\delta)A = -
(d\delta + \delta d) A = \delta d(\delta \Pi) = - \delta^2 d \Pi = 0$
since $\delta d \Pi = -d \delta \Pi$ from $\partial^2 \Pi = 0$.

Now, since our main purpose here is to exhibit the existence of the
new solutions we present only particular cases, leaving a complete
study for another publication. To show the existence of a stationary
solution $F_0$ ($\partial F_0 = 0$) relative to a given inertial
frame $I$ with adapted coordinates $\langle x^\mu\rangle = \langle
t,x^i\rangle$
introduce above, we choose the Hertz potential
\begin{equation}
{\Pi}_0 (t,\vec x) = \phi (\vec x) \exp{(\gamma^5\Omega t)}
\gamma^1\gamma^2 .
\end{equation}
Since $\partial^2 \Pi_0 =0$ we get that $\phi(\vec x)$ satisfies the
Helmholtz equation
\begin{equation}
\label{eq.12}
\nabla \phi + \Omega^2 \phi = 0
\end{equation}
The solutions of this equation are well known. Here we must comment
the fact that the scalar wave equation has solutions which travels
with speed less than $c$ is known since a long time, being
discovered by H. Bateman \cite{REF-8} in 1915 and rediscovered by
several people in the last few years, in particular by Barut \cite{REF-9}
\footnote{Barut found also subluminal
solutions of Maxwell equations, with a procedure different from the one
presented here, and
his solutions are also different from the above.}.  An elementary solution
of eq.(\ref{eq.12}) with spherical symmetry is
\begin{equation}
\label{eq.13}
\phi (\vec x) = C \frac{\sin{\Omega r}}{r}, \qquad r = \sqrt{x^2 + y^2 +
z^2} ,
\end{equation}
where $C$ is an arbitrary constant.
{}From this result we can write $F_0$ in polar coordinates $(r, \theta,
\varphi)$ as
\begin{eqnarray}
\label{eq.14}
&& F_0 = \frac{C}{r^3} [\sin \Omega t(\alpha \Omega r \sin \theta \sin
\varphi - \beta
\cos \theta\sin \theta \cos \varphi) \gamma^0 \gamma^1 \nonumber \\
&& - \sin \Omega t (\alpha\Omega r\sin\theta \cos \varphi + \beta
\sin\theta \cos\theta
\sin \varphi)\gamma^0 \gamma^2 \nonumber \\
&& + \sin \Omega t (\beta \sin^2 \theta - 2\alpha)\gamma^0 \gamma^3
+ \cos \Omega t
(\beta \sin^2 \theta - 2 \alpha ) \gamma^1 \gamma^2 \nonumber \\
&& + \cos\Omega t (\beta\sin\theta \cos\theta \sin\varphi +
\alpha\Omega r \sin \theta \cos
\varphi) \gamma^1 \gamma^3 \nonumber \\
&& + \cos \Omega t (-\beta \sin\theta \cos\theta \cos \varphi +
\alpha \Omega r \sin \theta
\sin\varphi)  \gamma^2 \gamma^3]
\end{eqnarray}
with $\alpha = \Omega r \cos \Omega r - \sin \Omega r$ and $\beta =
3\alpha + \Omega^2
r^2 \sin \Omega r$. Note that $F_0$ is {\it regular at the origin and
vanishes at infinity}.

Let us rewrite the above solution in terms of the old-fashioned
vector algebra. We have that
\begin{equation}
\vec{E}_0 = -\vec{W}\sin{\Omega t} , \quad
\vec{B}_0 = \vec{W}\cos{\Omega t} ,
\end{equation}
where we defined
\begin{equation}
\vec{W} = C \left( \frac{\alpha\Omega y}{r^3} - \frac{\beta x z}{r^5}, \quad
-\frac{\alpha\Omega x}{r^3} - \frac{\beta y z}{r^5} , \quad
\frac{\beta(x^2+y^2)}{r^5} - \frac{2\alpha}{r^3} \right) .
\end{equation}
One can explicitly verify that ${\rm div}\vec{W} = 0$, so that
${\rm div}\vec{E}_0 = {\rm div}\vec{B}_0 = 0$, and that
${\rm rot}\vec{W} + \Omega\vec{W} = 0$, so that
${\rm rot}\vec{E}_0 + \partial\vec{B}_0 /\partial t = 0$
and ${\rm rot}\vec{B}_0 - \partial\vec{E}_0 /\partial t = 0$.

We can show for a given $F$
\cite{REF-5} that $S^0 = {\frac{1}{2}} \widetilde F \gamma^0 F$
is the 1-form representing the energy density and the
Poynting vector.
We have
that $\vec E_0 \times \vec B_0 = 0$, so that there is
{\it no propagation\/} and the solution has vanishing angular momentum.
The
energy density is
\begin{equation}
u = \frac{1}{r^6} [\sin^2 \theta (\Omega^2 r^2 \alpha^2 + \cos^2
\theta \beta^2) +
(\beta \sin^2 \theta - 2\alpha)^2]
\end{equation}
Then $\int ud^3x =  \infty$. A finite energy solution can be
constructed by considering ``wave packets" with a distribution of
intrinsic frequencies $f(\Omega)$ \cite{REF-8}. These solutions will be
discussed in another publication. It is also very important to see
that $F^2_0 = F_0 \cdot F_0 + F_0 \wedge F_0 \neq 0$, i.e,
the field invariants are nonvanishing, differently of the usual
solutions $F_c$ of Maxwell equations that travel with constant speed
$c =1$
and for which $F^2_c = 0$.

To obtain a solution $F_<$ moving with velocity $0 < v < 1$ relative
to $I$ it is necessary only to make a Lorentz boost in the $x$
direction of the solution $F_0$.
Another way to obtain a solution $F_<'$ is
to get a new solution $\Phi_<(t, \vec x)$ of
the wave equation ($\partial^2 \Phi_< = 0$), like:
\begin{equation}
\label{eq.17}
\Phi_< (t, \vec x) = C \frac{\sin \Omega \xi_<}{\xi_<}
\exp [\gamma^5 (\omega t - k x)],
\end{equation}
\begin{equation}
\label{eq.18}
\omega^2 - k^2 = \Omega^2 ,
\end{equation}
with
$$
\xi_< = (\gamma^2_< (x-vt)^2 + y^2+z^2)^{1/2} ,
$$
$$
\gamma_< = \frac{1}{\sqrt{1-v^2}}, \quad  v = \frac{d\omega}{dk}
$$
Eq.(\ref{eq.18}) is a dispersion relation for a ``particle" moving with
velocity (here ``group velocity") $v_< = d\omega/dk < 1$.
{}From eq.(\ref{eq.17}) we
can  obtain a new solution $F_<'$ moving with $0 < v < 1$ and which
satisfies $\partial F_<' = 0$ by taking $F_<' =
\partial(\partial\cdot\Pi_<)$ with $\Pi_< = \Phi_< \gamma^1\gamma^2$.
The fact that $F_<'$ satisfies Maxwell
equations follows from the above theorem, but it has been explicitly
verified using REDUCE 3.5 \footnote{The use of REDUCE in Clifford
Algebras is discussed, for example, in \cite{REF-10,REF-10a}.}; unfortunately
the explicit form of $F_<'$
is very big to be given here.

Now, to obtain a superluminal solution of Maxwell equations it is
enough to observe that the function
\begin{equation}
\label{eq.19}
\Phi_> (t, \vec x) = C \frac{\sin \Omega \xi_>}{\xi_>} \exp [\gamma^5
(\omega t - k x)] ,
\end{equation}
\begin{equation}
\label{eq.20}
\omega^2 - k^2 = -\Omega^2 ,
\end{equation}
with
$$
\xi_> = (\gamma^2_> (x-v t)^2 - y^2 - z^2)^{1/2} ,
$$
$$
\gamma_> =
\frac{1}{\sqrt{v^2 -1}}, \quad v = \frac{d\omega}{dk} ,
$$
is a solution \cite{REF-11} of the wave equation $\partial^2 \Phi_> = 0$,
which travels with velocity $v = d\omega/dk > 1$. Writting $\Pi_> =
\Phi_> \gamma^1 \gamma^2$ we can obtain $F_> = \partial(\partial
\cdot \Pi_>)$ which
satisfies $\partial F_> = 0$. This is then a superluminal electromagnetic
configuration. The explicit form of $F_>$ is again very big to be
reproduced here, but again we verified it explicitly using REDUCE 3.5.
Also $F^2_< \neq 0$ which means that the field invariants are non
null. We will discuss more the properties of this and some others
extraordinary solutions of Maxwell equations in another paper.

\section{Subluminal and Superluminal solutions of the massless Dirac
equation}
In order to find these kinds of solutions of the massless
Dirac equation we shall make use of some ideas from supersymmetry.
Since a Dirac-Hestenes spinor field $\psi_\Sigma$ is given by
eq(\ref{eq.2}) we can
define a generalized potential for $\psi_\Sigma$. Indeed  for each
$\psi_\Sigma$ there exists  ${\cal A} = A+\gamma^5 B$, $A, B \in {\rm sec}
\bigwedge^1(M) \in {\rm sec}
{\cal C}\ell (M)$, such that
\begin{equation}
\label{eq.16}
\psi_\Sigma = \partial (A+ \gamma^5 B)
\end{equation}
The Dirac operator $\partial$ plays here a role analogous to that of
supersymmetry operator \cite{REF-12}. In fact, Clifford algebras
are $Z_2$-graded algebras, and ${\cal C}\ell^+{\cal C}\ell^+
\subset
{\cal C}\ell^+$, ${\cal C}\ell^\pm{\cal C}\ell^\mp \subset
{\cal C}\ell^{-}$, ${\cal C}\ell^{-}{\cal C}\ell^{-} \subset
{\cal C}\ell^+$, where ${\cal C}\ell^+$ [${\cal C}\ell^{-}$]
denotes the set of elements of ${\cal C}\ell$ with
even [odd] grade. Since the Dirac operator has vector
properties, its action transforms fields of even grade
into fields of odd grade, and vice-versa. Representing a
spinor field by means of nonhomogeneous forms of
even degree is therefore equivalent to find a bosonic
representation of a fermionic field.

The quantity ${\cal A}$ can be interpreted as a
kind of potential for the massless Dirac field.
Indeed, from eq.(\ref{eq.10}) with eq.(\ref{eq.16})
it follows that
\begin{equation}
\partial^2 {\cal A} = 0 ,
\end{equation}
or that
\begin{equation}
\partial^2 A = 0 , \qquad \partial^2 B = 0 .
\end{equation}

A simple subluminal solution at rest relative to the inertial frame $I =
e_0$ in the coordinates $\langle x^\mu\rangle$ naturally adapted to
$I$ is
\begin{equation}
{\cal A}_0 (t, \vec x) = \gamma^0 \phi (\vec x) \exp{(\gamma^5 \Omega t)}
\end{equation}
with $\phi (\vec x)$ given by eq.(\ref{eq.13}). We have
\begin{equation}
{\cal A}_0 = \frac{C}{r} (\sin \Omega r \cos \Omega t \gamma^0 - \sin
\Omega r \sin \Omega t
\gamma^1\gamma^2\gamma^3)
\end{equation}
Then
\begin{eqnarray}
&& \psi^0_\Sigma = \frac{C}{r^3} [
-\Omega r^2 \sin \Omega r \sin \Omega t  \nonumber \\
&& + \gamma^0\gamma^1 \lambda x \cos \Omega t
+ \gamma^0\gamma^2 \lambda y \cos \Omega t \nonumber \\
&& + \gamma^0\gamma^3 \lambda z \cos \Omega t
- \gamma^1\gamma^2 \lambda z \sin \Omega t \nonumber \\
&& + \gamma^1\gamma^3 \lambda y \sin \Omega t
- \gamma^2\gamma^3 \lambda x \sin \Omega t \nonumber \\
&& + \gamma^0\gamma^1\gamma^2\gamma^3 \Omega r^2 \sin \Omega r \cos
\Omega t ] ,
\end{eqnarray}
where
$\lambda = \Omega r \cos \Omega r - \sin \Omega r $.

The above solution in the usual formalism reads
\begin{equation}
\psi^0 = \left( \begin{array}{c}
\displaystyle{i\sin{\Omega t}\left(\frac{\lambda z}{r^3} +
i\frac{\Omega}{r}\sin{\Omega r}\right)} \\
\displaystyle{i\sin{\Omega t} \left( \frac{x+i y}{r^3}\right)\lambda} \\
\displaystyle{-\cos{\Omega t}\left(\frac{\lambda z}{r^3} +
i\frac{\Omega}{r}\sin{\Omega r}\right)} \\
\displaystyle{-\cos{\Omega t}\left( \frac{x+i y}{r^3}\right)\lambda}
\end{array} \right)
\end{equation}
and one can explicitly verify that indeed $\partial \psi^0 = 0$.

Other subluminal solutions can be obtainned by appropriated
boosts.
An explicit superluminal solution $\psi^>_\Sigma$ can be obtained by
writting
\begin{equation}
{\cal A}_> (t, \vec x) = \gamma^0 \Phi_> (t, \vec x)
\end{equation}
with $\Phi_>$ given by eq.(\ref{eq.19}) and $\psi^>_\Sigma =
[\partial \Phi_>
(t, \vec x)]\gamma^0$. Again the explicit form of $\Psi^>_\Sigma$ is
very big and
will be not written  here, but it has been verified using REDUCE.
Barut \cite{REF-12A} has found subluminal solutions of
$\partial \Psi = 0$, where $\Psi$ is the usual Dirac spinor
field,  with a different method from the one used here; our one is much
simple since it is representation free and uses elegant tools from
supersymmetry.

\section{Conclusions}

We want to discuss three possible implications of the
results we have shown.

\vspace{1ex}

\noindent {\bf (i)} If the superluminal solutions of at least Maxwell equations
are realized in Nature we can have a breakdown of Lorentz invariance.
Indeed, suppose $I = \partial/\partial t$ is the fundamental
reference frame and $I^{\prime} = (1/\sqrt{1-V^2})\partial/\partial t -
V/\sqrt{1-V^2}\partial/\partial x$ is the laboratory frame
(an inertial frame). Suppose that $F_>$ is a superluminal
solution of Maxwell equations, i.e., $\partial F_> = 0$
($\omega^\prime{}^2 - k^{\prime}{}^2 = -\Omega^2$), travelling
forward in time according to $I$. Then, the validity of
active Lorentz invariance implies that there exists $R \in
{\rm Spin}_{+}(1,3)$ such that $F_>^{\prime} = RF_> \tilde{R}$
satisfies $\partial F_>^{\prime} = 0$ with $F_>^{\prime}$
going backward in time (and carrying negative energy)
according to $I$. This solution can be interpreted as an
``anti-field'' coming from the past, and is a good solution.
However, the physical equivalence of all inertial reference
frames implies that according to $I^{\prime}$ there exists
solutions $F^{\prime\prime}_>$ of Maxwell equations travelling
forward in time and carrying positive energy $E^{\prime} =
\omega^\prime$ ($\hbar = 1$) according to $I^\prime$ but
travelling backward in time (and carrying negative
energy) according to $I$. The field $F^{\prime\prime}_>$
can be absorved, e.g., by a detector in periodic motion in $I$ (it is
enough that at the time of absortion the detector has
relative to $I$ the velocity $V$ of the $I^{\prime}$
frame). This generates as is well known a causal
paradox \cite{REF-16} (Tolman-Regge paradox). The possible
solution is to say that $I$ and $I^{\prime}$ are not
physically equivalent. We then have the following:
$I^\prime$ cannot send to some observers (integral lines)
of the $I$ reference frame a superluminal signal
such that $\omega^\prime < (V/\sqrt{1-V^2})\Omega$.
When $\omega^\prime = (V/\sqrt{1-V^2})\Omega$ the
superluminal generator of $I^\prime$ stops working
for $k^\prime$ in  some spacetime directions, and
an observer in $I$ can calculate his absolute velocity which
is $V = \omega^\prime /\sqrt{\omega^\prime{}^2 + \Omega^\prime{}^2}$.
We must also call the reader's attention that recently
Nimtz \cite{REF-13,REF-13a}
transmited Mozart's symphony \# 40 at 4.7 $c$ through a
retangular wave guide, that as is now well known \cite{REF-14} acts
like a potential barrier for light. Important related results have
also been obtained in \cite{REF-15}. We can show easily that under
Nimtz experimental conditions the solution of Maxwell equations in the
guide gives a dispersion relation like eq(\ref{eq.20}), i.e,
corresponding to superluminal propagation \cite{REF-16}.
We shall discuss this
issue in details elsewhere.

\vspace{1ex}
\noindent {\bf (ii)} The existence of the subluminal solutions $F_<$ are very
important for the following reason: Recently \cite{REF-17,REF-17a} we proved
that $\partial F =0$ for $F^2\neq 0$ is equivalent under certain
conditions to a Dirac-Hestenes
equation $\partial \psi_\Sigma \gamma^1\gamma^2 + m\psi_\Sigma
\gamma^\circ =0$, where
$F = \psi_\Sigma \gamma^1\gamma^2 \widetilde{\psi}_\Sigma$.
This means that eventually particles are special stationary
electromagnetic waves and a de Broglie interpretation of quantum
mechanics seems possible \cite{REF-18}. We will discuss this issue in
details elsewhere.

\vspace{1ex}
\noindent {\bf (iii)} Finally the existence of subluminal and  superluminal
solutions for
$\partial \psi_\Sigma =0$ (which reduces to Weyl equation
for $\psi_\Sigma$
a Weyl spinor) may be
important to solve some of the mysteries associated with neutrinos.
Indeed if neutrinos can be produced in the subluminal and
superluminal modes
-- see \cite{REF-19,REF-19a} for some experimental evidences for superluminal
neutrinos -- then they can eventually escape detection on earth after
leaving the sun. Moreover, for neutrinos in a subluminal mode
it would be possible to define a kind of ``effective mass''.
Recently some cosmological evidences that
neutrinos may have a nonvanishing mass have been discussed \cite{REF-20}.
One such ``effective mass'' could be responsible
for those cosmological evidences,
and in such a way that
we can still have a left-handed neutrino since it would satisfies
the Weyl equation.
We are going to study this proposal in a forthcoming
paper.

\acknowledgements{We are grateful to CNPq and FAEP-UNICAMP for the finnancial
support. The authors are grateful to the members of the Mathematical
Physics Group of IMECC-UNICAMP and to B.A.R. Ferrari for many usefull
discussions. One of the authors (J.V.) wishes to thank J. Keller
and A. Rodriguez for their very kind hospitality at FESC - UNAM and
UASLP, respectively.}

\vspace{5ex}

\noindent {\bf Note added in proof.} After we finished this
paper we have been informed by Professor Ziolkowski that he
and collaborators found also superluminal solutions
of the scalar wave equation and also of Maxwell equations
and even Klein-Gordon equation \cite{REF-21,REF-21a,REF-21b}.
Also Dr. Lu and collaborators found a very interesting
``superluminal'' solution of the scalar wave equation
\cite{REF-22,REF-22a} and Lu even realized an approximation for
his solution (the so called X waves) as a nondispersive
pressure wave in water which travels with velocity
1.002 $c$, where $c$ here is the velocity of sound in water!

The solutions found by Lu can be used to construct Hertz
potentials for the Maxwell equations and then to generate
superluminal electromagnetic field configurations. We believe
that it is in principle possible to build such fields
with appropriate devices. This is also the opinion of
Dr. Lu. Also some of the Ziolkowski solutions,
according to his opinion, may be realized
in the physical world. We shall
discuss these points in another opportunity.

\end{document}